\documentclass[aps,prb,twocolumn,superscriptaddress]{revtex4-1}
\usepackage{graphicx}
\usepackage{bm}
\usepackage{amsmath}
\usepackage{amssymb}
\usepackage{euscript}
\usepackage{verbatim}
\usepackage{setspace}
\usepackage{xcolor}
\usepackage{amsfonts}
\usepackage{braket}
\usepackage{subfigure}

\newcommand{\be}{\begin{equation}}
\newcommand{\ee}{\end{equation}}
\newcommand{\bea}{\begin{eqnarray}}
\newcommand{\eea}{\end{eqnarray}}

\begin{document}

\title{Transport through dynamic pseudo-gauge fields and snake states 
	in a   Corbino  
	geometry}
\author{Jonathan Amasay}
\affiliation{School of Physics and Astronomy, Tel Aviv University, Tel Aviv 6997801, Israel}

\author{Eran Sela}
\affiliation{School of Physics and Astronomy, Tel Aviv University, Tel Aviv 6997801, Israel}

\begin{abstract}
	We propose a Corbino-disk geometry of a graphene membrane under  out-of-plane strain deformations as a convenient path to detect pseudo-magnetic and electric fields via  electronic transport.
		The three-fold symmetric pseudo-magnetic field changes sign six times as function of angle, leading to snake states connecting the inner and outer contacts and to nearly quantized transport.
		For dynamical strain obtained upon AC gating, the system supports an AC pseudo-electric field which, in the presence of the pseudo-magnetic field, produces a net electronic charge current in the absence of an external voltage, via a  pseudo-Hall effect. 
\end{abstract}

\maketitle	

\linespread{1} 

\section{Introduction}
\begin{figure}[h]
	\includegraphics[width=0.5\textwidth]{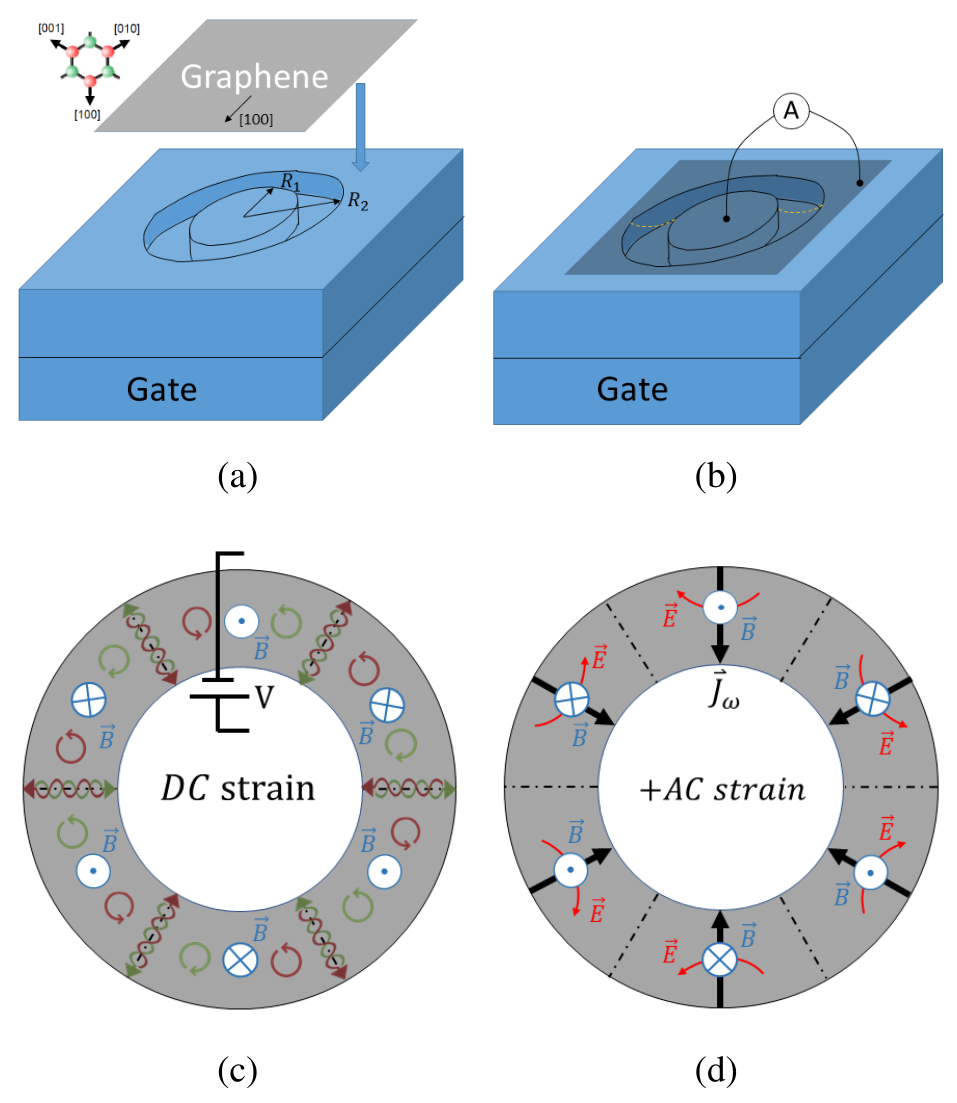}
	\caption{Schematics of our system. (a) Graphene membrane deposited on a circular hole and stretched via a gate. (b) Current is measured between the inner and outer contacts. (c) Three-fold symmetric pseudo-magnetic field  which changes sign at angles $\frac{2 \pi}{6}j$ with integer $j$. At these angles, corresponding to crystallographic zigzag directions, snake states appear and carry the current upon applying an external voltage $V$. We denote trajectories of valley $K$ and $K'$ in green and red, respectively. (d) A pseudo-electric field   is generated by an additional  dynamic strain, resulting in an AC pseudo-Hall current  (black arrows) of the same direction for valleys $K$ and $K'$.  
	}
	\label{system}
\end{figure}

Large pseudo-magnetic fields 
have been detected in the form of quantized pseudo-Landau levels in graphene samples with macroscopic strain deformations, using scanning tunneling microscopy~\cite{levy2010strain,zhu2014pseudomagnetic,jiang2017visualizing} and angular resolved photo emission~\cite{nigge2019room}. The pseudo-magnetic field, $\vec{B}=\vec{\nabla} \times \vec{A}$, results from strain gradients, which create an effective space dependent vector potential on the Dirac electrons~\cite{guinea2008gauge,vozmediano2010gauge,masir2013pseudo,Si2016strain,Deji2017review}. Microscopically, this effect results from the modification  of the  hopping  amplitudes  by  strain~\cite{guinea2010energy,vozmediano2010gauge}. The  pseudo-magnetic field leads to 
electronic transport properties akin to the quantum Hall regime, while time-reversal symmetry is preserved~\cite{fogler2008pseudomagnetic,low2010strain,fujita2010valley,milovanovic2016strained,yesilyurt2016perfect,settnes2017valley,soto2018electronic,bhagat2019pseudo,liu2020strain,wu2018quantum}. 
Various proposals exist for special geometries leading to uniform pseudo-magnetic fields~\cite{guinea2010energy,guinea2010generating,chaves2010wave,gomes2012designer,shioya2014straining,zhu2015programmable,verbiest2015uniformity,downs2016towards,sela2020quantum}.
Yet, finding systems showing unambiguous signatures of pseudo-fields on electronic transport had remained challenging.

Dynamic strain, e.g. produced by lattice vibrations,  leads to pseudo-electric fields, $\vec{E}=-\frac{\partial \vec{A}}{\partial t}$, and consequently to valley-currents and phonon damping~\cite{von2009synthetic}.
These currents, as the pseudo-vector potential $\vec{A}$, have opposite signs in the two valleys in the Brillouin zone, hence they do not carry charge.
When the strain field has both spatial gradients and time dependence, however, the resulting pseudo-magnetic and electric fields coexist, and thereby produce charge currents due to a Hall-drift velocity $\vec{v}_d=[(\vec{B}\times \vec{E})/|B|^2]$ ~\cite{sela2020quantum,sabsovich2021helical}. 
Previous suggestions to observe this pseudo-Hall electro-mechanical response, converting strain into electric current, considered stretched graphene ribbons of trapezoidal shape~\cite{zhu2015programmable, sela2020quantum}, thereby creating strain gradients. 
Such devices have been recently fabricated~\cite{sonntag2019engineering}. Upon adding an AC component to the strain, e.g. piezoelectrically, one creates pseudoelectic fields on top of pseudo-magnetic fields. The vector product of the pseudo $B$- and $E$-fields produces a Hall drift charge current. An experimental disadvantage of this
geometry is its opens ends, at which large strains~\cite{zhu2015programmable, sela2020quantum} can lead to rapture.

In this work we propose an alternative geometry to demonstrate electronic transport  via pseudo-magnetic and electric fields, based on a  Corbino-disk  geometry. While this setup was proposed in a modified geometry allowing to tailor the profile of the pseudo-magnetic field by shape design~\cite{jones2017quantized}, here we consider a circularly symmetric geometry.  As shown in Fig.~\ref{system}(a,b) we consider a 
graphene membrane deposited on a ring-shaped hole, separating the inner and outer metallic contacts. The membrane is  then strained out-of-plane via a gate capacitor.
The electric currents resulting either by an external voltage in the presence of a static strain [Fig.~\ref{system}(c)] or from the dynamic strain [Fig.~\ref{system}(d)] can be detected between the inner and outer contacts. This system can be viewed, from the viewpoint of elasticity, as a periodic boundary condition version of the ribbon with varying width~\cite{zhu2015programmable,sela2020quantum}, avoiding open ends and allowing larger oscillation frequencies.

In polar coordinates, the pseudo-vector potential is given by~\cite{guinea2010energy} 
\bea
A_r&=&C \left( (u_{rr}-u_{\theta \theta})\cos 3 \theta - 2 u_{r \theta} \sin 3 \theta \right), \nonumber \\
A_\theta&=&C \left( (-u_{rr}+u_{\theta \theta})\sin 3 \theta + 2 u_{r \theta} \cos 3 \theta \right),
\label{vectorP}
\eea
where $C=\frac{\beta t}{e v_F}=6.25 \rm{T} \cdot \rm{\mu m}$, $\beta=-\frac{\partial \ln  t }{\partial \ln a }\approx 2.5$ (with $a$ the lattice constant), $t\approx 2.5eV$ is the hopping parameter, $e$ is the electron charge, and $v_F=10^{6}\rm{m/s}$ is the Fermi velocity in graphene. The angular dependence reflects the three-fold symmetry of the honeycomb lattice, where $\theta=0$ corresponds to a zigzag direction. For the circularly symmetric geometry in Fig.~\ref{system}, including an  out-of-plane displacement $h$, the strain components are~\cite{timoshenko1959theory}
\bea
u_{rr}=\frac{\partial u_r}{\partial r }+\frac{1}{2} \left(\frac{\partial h}{\partial r}\right)^2,~~~~u_{\theta \theta}=\frac{u_r}{r},  ~~~~u_{r \theta}=0.
\label{polarstraintensor}
\eea
The pseudo-magnetic field $B=(\vec{\nabla} \times \vec{A})_z=-\frac{1}{r}\frac{\partial A_r}{\partial \theta}+\frac{\partial A_\theta}{\partial r}+\frac{A_\theta}{r}$, can then be written as
\bea
B(r,\theta)&=&C \mathcal{B}(r)\sin 3 \theta,~~~~\mathcal{B}(r)=-\frac{\partial \mathcal{U} \left( r \right)}{\partial r}+\frac{2}{r}\mathcal{U} \left( r \right),\nonumber \\
\mathcal{U} &=&u_{rr}-u_{\theta \theta}.
\label{magneticfield}
\eea
As for the pseudo-electric field $\vec{E}=-\frac{\partial \vec{A}}{\partial t}$, we have 
\bea
(E_r,E_\theta)=C \partial_t \mathcal{U}(r)(\cos 3\theta,- \sin 3\theta).
\label{electricfield}
\eea
The explicit form of the deformation fields $u_r$ and $h$ will be determined below. Here, we observe from the  relation $B \propto \sin 3 \theta$, that $B$ changes sign at specific angles, $\theta = \frac{2\pi}{6} j$, with integer $j$ (see e.g. Refs.~\onlinecite{moldovan2013electronic,carrillo2018enhanced}). These are the zigzag directions. This leads to snake states~\cite{oroszlany2008theory,ghosh2008conductance,jones2017quantized}, becoming quantized edge states separated by gapped localized states  at large magnetic field, see Fig.~\ref{system}(c). Indeed, we show that for reasonable gating and device dimensions, $\sim 1~$Tesla pseudo-magnetic fields are produced. Once these snake states will connect the inner and outer transport contacts, this system will feature quantized conductance~\cite{oroszlany2008theory,ghosh2008conductance,jones2017quantized}.

Now consider the dynamic regime, with an extra small AC pseudo-electric field, see Fig.~\ref{system}(d). The static magnetic field $\propto \sin 3 \theta$, as well as the azimuthal AC pseudo-electric field $\propto \sin 3 \theta$ average spatially to zero. Their product, however, produces a drift pseudo-Hall response $\propto  \vec{B}\times \vec{E}$ resulting in a net radial current. Thus the Corbino geometry allows to observe the AC pseudo-Hall effect~\cite{sela2020quantum} although the pseudo-fields are nonuniform.

The paper is organized as follows. In Sec.~\ref{sec:model} we present the elastic model and solve it for the pseudo-$B$ and -$E$ fields, along with the displacement and height fields. We obtain analytical expressions in the narrow ring limit.
Then, in Sec.~\ref{sec:response} we study the electric conductance of the static membrane under an external voltage difference, while in Sec.~\ref{sec:responseAC} we focus on the dynamic case where the pseudo-magnetic and electric fields coexist and lead to a charge current. In both cases we discuss the classical versus quantized regimes. In Sec.~\ref{appendix2} we analyze the snake states as a solutions of the Dirac equation in a periodically vaying magnetic field.  We conclude in  Sec.~\ref{sec:conclusions}. 

\section{Elastic Model}
\label{sec:model}

As shown in Fig.~\ref{system}, we consider a graphene membrane suspended over a ring shaped hole with radii $R_1<R_2$, equivalent to a drum with a central circular fixed region. We envision a control of the out-of-plane deformation by an electric gate capacitor.

The total mechanical energy of the membrane is given by $E_{tot}=E_{elas}+E_{bend}+E_{field}$. The elastic energy is given by $E_{elas}=\int d^{2}r\left(\frac{\lambda}{2}\left(\sum_{i}u_{ii}\right)^{2}+\mu\sum_{i,j}u_{ij}^{2}\right)$, where $u_{ij}$ is the strain tensor, and $\lambda\approx 2.4 eV {\AA}^{-2}$ and  $\mu\approx 9.9 eV {\AA}^{-2}$ are the elastic Lam\`{e} coefficints for graphene. The bending energy is given by $E_{bend}=\frac{\kappa}{2}\int d^{2}r\left(\nabla^{2}h \right)^{2}$, where $\kappa \approx 1eV$ is the bending rigidity. $E_{bend}$ is negligible with respect to the elastic energy for large enough out-of-plane deformation~\cite{vozmediano2010gauge}, $h \gtrsim
\sqrt{\kappa/{\rm{max}}\left(\lambda,\mu \right)}\approx3{\AA}$. The term $E_{field}$ describes the electric field from a gate that exerts pressure on the membrane, $E_{field}=-\int d^{2}r P h$, where $P=\frac{2\pi e^{2}n^{2}}{\epsilon}$ is dictated by the electron density in the membrane, $n$.

By circular symmetry, the strain tensor has no $\theta$ dependence, and there is no tangential displacement, $u_{\theta}=0$. Transforming into polar coordinates and integrating over $\theta$ yields
\begin{align}
E_{tot}&=2\pi  \int rdr \left(\frac{\lambda}{2}\left(u_{rr}+\frac{u_r}{r}\right)^{2}+ \mu\left(u_{rr}^{2} +\left(\frac{u_r}{r}\right)^{2}\right)\right)\nonumber\\
&-2\pi\int rdr Ph.
\label{polarenergy}
\end{align}
As we can see in Eqs.~(\ref{polarstraintensor}) and~(\ref{polarenergy}), the total energy is a functional of the radial displacement $u_r(r)$ and the height function $h(r)$. Minimizing, we obtain the pair of equations~\cite{timoshenko1959theory}
\begin{align}
\frac{\partial^{2}u_{r}}{\partial^{2}r}=&-\frac{1}{r}\frac{\partial u_{r}}{\partial r}+\frac{u_{r}}{r^{2}}\nonumber\\
&-\left(\frac{\partial h}{\partial r}\right)\left(\frac{\partial^{2}h}{\partial r^{2}}\right)-\left(\frac{\mu}{\lambda+2\mu}\right)\frac{1}{r}\left(\frac{\partial h}{\partial r}\right)^{2},\nonumber
\end{align}
\begin{align}
&\frac{\partial}{\partial r}\left(r\left(\frac{\lambda}{2}+\mu\right)\left(\left(\frac{\partial h}{\partial r}\right)^{3}+2\frac{\partial u_{r}}{\partial r}\frac{\partial h}{\partial r}\right)+\lambda u_{r}\frac{\partial h}{\partial r}\right)\nonumber\\
&+Pr=0,
\label{diffuandh}
\end{align}
supplemented by the boundary conditions $u_r\left(R_1\right)=u_r\left(R_2\right)=h\left(R_1\right)=h\left(R_2\right)=0$. 
Before proceeding with a numerical solution, 
we treat analytically the limit of a narrow ring.

\subsection{Narrow ring limit $(\Delta R \ll R)$}

We denote the width of the ring as $\Delta R=R_2-R_1$, and its avarage radius $R=\frac{R_1+R_2}{2}$. Now  consider the limit $\Delta R \ll R$ and define a new radial variable, $\rho=r-R$, $\rho \in [\frac{-\Delta R}{2},\frac{\Delta R}{2}]$, in terms of which, up to zero order in $\frac{\Delta R}{R}$, Eqs.~(\ref{diffuandh}) become
\bea
u''+h'h''&=&0,\nonumber \\
h''\left(h'^{2}+2u'\right)+\frac{2P}{\lambda+2\mu}&=&0.
\label{diffuandhlim}
\eea
We denoted $u \equiv u_r$.
One can solve this equation and obtain the profile of in- and out-of-plane deformations $u$ and $h$, 
\bea
&h(\rho)=h_{0}\left(1-\left(\frac{2\rho}{\Delta R}\right)^{2}\right),\nonumber\\
\nonumber\\
&u(\rho)=\frac{32}{3}\left(\frac{h_{0}}{\Delta R^2}\right)^{2}\left(\left(\frac{\Delta R}{2}\right)^{2}\rho-\rho^{3} \right),
\label{limitsol}
\eea
with maximal out-of-plane displacement 
\be
h_{0}=\left( \frac{\Delta R}{2} \right)^{2}\left(\frac{3P}{\left(\lambda +2\mu \right)\Delta R^{2}}\right)^{\frac{1}{3}},
\ee
which scales with the electron density as~\cite{vozmediano2010gauge} $h_{0}\propto n^{\frac{2}{3}}$.

\begin{figure*}
	\includegraphics[width=0.95\textwidth]{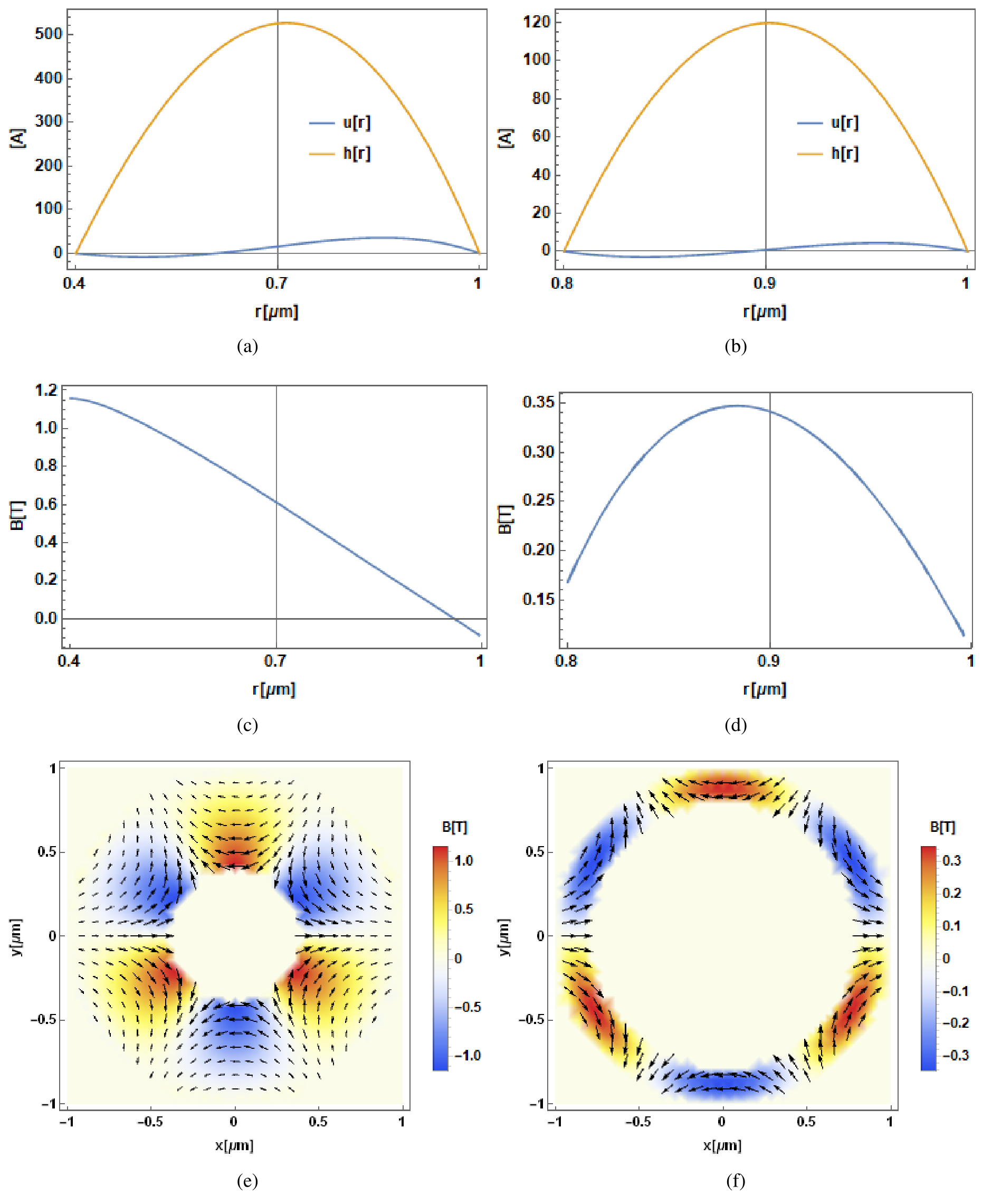}
	\caption{Numerical results for two different geometries of a wide ring $R_1=0.4\mu m$, and $R_2=1\mu m$ in the left panels and a narrower ring $R_1=0.8\mu m$, and $R_2=1\mu m$ in the right panels. We compute the out-of-plane $h(r)$ and in-plane $u(r)$ deformation profiles (a,b); the radial dependence of the pseudo-magnetic field $\mathcal{B}(r)$ (c,d); and a color map of the pseudo-magnetic field along with stream lines of the vector potential $\vec{A}$ (e,f).}
	\label{ringg}
\end{figure*}

The pseudo-magnetic field is then obtained from Eq.~(\ref{magneticfield}),
\bea
\mathcal{B}=\left(-\frac{\partial}{\partial r} +\frac{2}{r}\right)\left( u'+\frac{1}{2}h'^2 -\frac{u}{r}\right).
\eea 
We can see from Eq.~(\ref{limitsol}) that it vanishes to zero order in $\frac{1}{R}$. Including the leading expansion in $\Delta R/R$ to Eq.~(\ref{diffuandhlim}) we obtain (see appendix \ref{appendix1})
\be
\label{Banalitic}
\mathcal{B}=\frac{32h_{0}^{2}}{3\Delta R^{2}R}\left(1-6 \left( \frac{\lambda +\mu}{\lambda+2\mu} \right) \left( \frac{\rho}{\Delta R} \right)^{2} \right).
\ee
This is the main result of this section. It strictly applies in the limit $\Delta R \ll R$, but it is serves as a useful  analytic approximation away from from this limit.
By mirror $h \to -h$ symmetry the magnetic field scales quadratically with $h_0$. It also has a parabolic dependence on $\rho$. It does not change sign with $\rho$ as long as $\mu>\lambda$, and the ratio between its maximal and minimal values is $\frac{B_{max}}{B_{min}}=4 \frac{\mu+\lambda/2}{\mu-\lambda} \cong 5.9$. 

\subsection{Numerical solution}
Now we present numerical solutions of the elasticity equations (\ref{diffuandh}). 
Numerical simulations near the narrow ring limit perfectly agree with the analytic results in the previous subsection. This is shown for  $R_{1}=10\mu {\rm{m}}$ and $R_{2}=10.02 \mu {\rm{m}}$ in appendix \ref{appendix1}.

 We now concentrate on more realistic membrane dimensions with $R_2=1\mu {\rm{m}}$, and either $R_1=0.4\mu {\rm{m}}$ (left panels of Fig.~\ref{ringg}), or $R_1=0.8\mu {\rm{m}}$  (right panels of Fig.~\ref{ringg}). In both cases the electron density is  $n=7.5\cdot 10^{12}~{\rm{cm}}^{-2}$. We can see that the height function is still nearly parabolic as in the narrow ring limit, see Eq.~\ref{limitsol}. 
 However, while in the narrow width limit the in-plane displacement $u$ changes sign at $(R_1+R_2)/2$, i.e. at the center of the ring, away from the narrow ring limit this point shifts towards smaller radii. Similarly,  in the narrow ring limit the pseudo-magnetic field has a maximum at $(R_1+R_2)/2$, which shifts towards smaller radii away from this limit, and eventually we encounter situations where $B$ changes sign as function of $r$ (see Fig.~\ref{ringg}(c)). We can avoid such situations by considering situations closer to the narrow ring limit.
We plot stream lines of the vector potential Eq.~(\ref{vectorP}) which indicate the direction of the pseudo-electric field which would result from an oscillating strain. 

\section{Electric Response - static case}
\label{sec:response}
In the previous section we determined the pseudo-fields in the srtained Corbino-disk geometry. Next we discuss their effect on electronic transport. In this section we consider the static case as in Fig.~\ref{system}(c), applying an external voltage difference between the contacts. The dynamic pseudo-Hall effect is discussed in Sec.~\ref{sec:responseAC}. In both static and dynamic regimes, we separate our treatment to the small magnetic field regime described by a Drude theory, and the large magnetic field quantum-Hall regime described by edge state transport, which we describe in more detail in Sec.~\ref{appendix2}.

\subsection{Classical theory}
\label{se:classical}
We first consider the classical equations for the currents $j_\pm$ and  densities $n_\pm$ in the two valleys~\cite{sela2020quantum}
\bea
\vec{j}_{\pm}= \sigma \left(\vec{E}_{ext} \pm \vec{E}(r,t) \right) -D\vec{\nabla}n_{\pm}\mp   \omega_{c}  \tau  \vec{j}_{\pm}\times\hat{z},
\label{calssiccurrent}
\eea
together with their  conservation laws, $\vec{\nabla} \cdot \vec{j} _\pm +\partial_t n_\pm=0$. We ignore intervalley scattering by assuming variations of effective potentials occurring on scales exceeding the atomic scale. 
Here $\sigma$ is the electrical conductivity per spin and per valley, $D$ is the diffusion coefficient, $\tau$ is the relaxation time and $\omega_c$ is the (space dependent) cyclotron frequency $\propto B$. In this section there is no pseudo-electric field $\vec{E}(r,t) =0$, and $\vec{E}_{ext} $ is applied radially by the external voltage $V$ as in Fig.~\ref{system}(c).

For simplicity we (i) consider the narrow ring limit $\Delta R \ll R$ and (ii) ignore the $r$ dependence of the electromagnetic fields in Eqs.~(\ref{electricfield}),~(\ref{magneticfield}), but we do consider their $\theta$-dependence. This allows to capture the $\theta$-dependence of the currents, 
\bea
j_{r,\pm}&=&\sigma E_{ext} \mp (\omega_c \tau) j_\theta \sin 3 \theta, \nonumber \\
j_{\theta,\pm}&=&\pm (\omega_c \tau) j_r \sin 3 \theta, 
\eea
Here,  $E_{ext}=V/\Delta R$ is determined by the external voltage. The diffusion term, describing screening, will play a crucial role in Sec.~\ref{sec:responseAC} in the case of an internal pseudo-electric field, and is ignored here since the external voltage cannot be screened by the graphene membrane. This leads to a radial charge current density 
$j_r= j_{r+}+j_{r-}=2 \frac{\sigma E_{ext}}{1+(\omega_c \tau)^2 \sin^2 3 \theta}$, which is maximal at the angles  where $B=0$, i.e. $\theta=\frac{2\pi}{6} j$, with integer $j$. 
The total radial current (including spin) is given by 
\be
\frac{I}{V} =4 \sigma \frac{R}{\Delta R}  \int_0^{2\pi} d\theta  \frac{1}{1+(\omega_c \tau)^2 \sin^2 3 \theta}.
\ee
In the limit $\omega_c \tau \ll 1$, the pseudo-fields lead to a small modification with respect to the $B=0$ Ohm's law, $\frac{I}{V} = 4 \sigma \frac{2\pi R}{\Delta R}$.

\subsection{Edge state transport}
\label{se:quantum}
In the large $B$  pseudo-quantum Hall regime, $\omega_c \tau \gg 1$, the  wave functions become localized in the bulk and the classical Eq.~(\ref{calssiccurrent}) is not valid. 
In our geometry the magnetic field 
switches sign along 6 zigzag directions.
We assume that each one of the 6 wedges with fixed ${\rm{sign}} (B)$  stabilizes a quantum Hall state with a well defined  filling factor $\pm \nu$, surrounded by  edge (snake) states. When the electronic contacts directly connect to these edge states, as in Fig.~1(c),
we expect a quantized current,
\be
\label{6Landauer}
\frac{I}{V} = 6 \nu \frac{e^2}{h},
\ee
see Sec.~\ref{appendix2} for a derivation.

The conductance quantization may be spoiled  by
various effects. Consider first the role of the contacts~\cite{bahamon2013effective}. As can be  seen in Fig.~\ref{ringg}~(e,f), see also Eq.~(\ref{Banalitic}), the pseudomagnetic field weakens near the contacts,  
 hence a finite diffusive region may appear between  the external contacts and the edge states, and affect the conductance. Note that for special non-circular geometries the magnetic field profile can be made more piecewise uniform~\cite{jones2017quantized}. The quantization of the conductance also requires the absence of intervalley scattering, which
 may unavoidably appear near the contacts.  On the other hand, the conductance is still expected to show step-like dependence on the number of connecting edge (snake) states depending on $\nu$. 

\section{Dynamic pseudo-Hall effect}
\label{sec:responseAC}
We now consider a small dynamic strain component on top of the static strain. This can be achieved by  an AC gate modulation on top of a DC value. We turn off the external source-drain voltage. Explicitly, using Eqs.~(\ref{magneticfield}) and (\ref{electricfield}), with a strain contribution $\mathcal{U} = \mathcal{U}^{DC} + \mathcal{U}^{AC} e^{i \omega t}$ where $\mathcal{U}^{AC} \ll \mathcal{U}^{DC}$, we have
\bea
B&=&C\mathcal{B}(r) \sin 3 \theta, \nonumber \\
(E_r,E_\theta)&=&E_\omega(\cos 3 \theta,-\sin 3 \theta) e^{i \omega t},
\eea
where $\mathcal{B}(r) \approx -\frac{\partial \mathcal{U}^{DC} }{\partial r}+\frac{2}{r}\mathcal{U}^{DC} $, and $E_\omega =C i \omega \mathcal{U}^{AC} $. We ignore the weak AC component of $B$.

Denoting $j \equiv j_+$, and remembering that $j_-$ is obtained by flipping the signs of $E_\omega$ and $\omega_c$, we obtain
\begin{align}
\label{firsteq}
&j_{r}= \sigma E_\omega \cos 3\theta - \left( \omega_{c} \tau\right) j_{\theta}\sin 3\theta ,
\nonumber\\
&j_{\theta}= \sigma  E_\omega \sin 3\theta -iD\frac{\partial_{\theta}^{2}j_{\theta}}{\omega R^{2}} + \left( \omega_{c} \tau\right) j_{r}\sin 3\theta,
\end{align}
yielding
\begin{align}
&i\frac{D}{\omega R^{2}}\partial_{\theta}^{2}j_{\theta}+j_{\theta}\left( 1+\left( \omega_{c} \tau\right)^{2}\sin^{2} 3\theta \right)\nonumber\\
&- \sigma E_\omega \sin 3\theta \left(1 + \omega_{c} \tau \cos 3 \theta  \right)=0.
\label{jt}
\end{align}
The parameter $\frac{D}{\omega R^2}$ describes to what extent the pseudo-electric field is screened by a rearrangement of the electronic density~\cite{sela2020quantum}. In the low frequency limit $\frac{D}{\omega R^2}\gg 1$ the electrons have sufficient time to diffuse thus screening the pseudo-electric field. The opposite is true in the  high frequency regime $\frac{D}{\omega R^2}\ll 1$.

\begin{figure}[h]
	\includegraphics[width=0.4\textwidth]{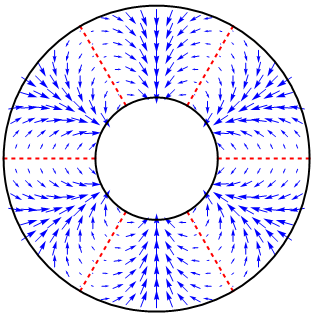}
	\caption{Charge current density in the classical high frequency regime as given in Eq.~(\ref{ineffectivescreening}). The red dashed lines represent the angles where the magnetic field vanishes.
	}
	\label{jtot}
\end{figure}

Consider the ineffective screening, high frequency regime $\frac{D}{\omega R^2}\ll 1$. 
Solving Eq.~(\ref{jt}) we obtain the angular dependence of the classical current components,
\bea
\label{ineffectivescreening}
\frac{j_{\theta}}{\sigma E_\omega}&=&   \frac{ \sin 3\theta \left( 1 + \left( \omega_{c} \tau\right) \cos 3\theta \right)}{ 1+\left( \omega_{c} \tau\right)^{2} \sin^{2} 3\theta},  \\
\frac{j_{r}}{\sigma E_\omega}&=&  \cos 3 \theta - \frac{\omega_c \tau \sin^2 3\theta \left( 1+ \left( \omega_{c} \tau\right) \cos 3\theta \right)}{ 1+\left( \omega_{c} \tau\right)^{2} \sin^{2} 3\theta}. \nonumber
\eea
The resulting charge current density $j_++j_-$ in the first half of the period  is plotted in Fig.~\ref{jtot}. 
The radial current contains a valley-even component with a finite angular average,
\bea
\label{eq:19}
I_{\omega^{>}} =  4\sigma E_\omega R  \int d \theta   \frac{\omega_c \tau \sin^2 3\theta }{1+(\omega_c \tau)^2 \sin^2 3\theta}.  
\eea
This is an AC current flowing at the same frequency as the pseudo-electric field. As seen in Fig.~\ref{jtot}, it is maximal in the regions where $|B(\theta)|$ is maximal.

In the opposite strong screening, low frequency regime $\frac{D}{\omega R^2}\gg 1$, also assuming that $\omega_c \tau \ll 1$, Eq.~(\ref{jt}) gives
\be
j_\theta=i \frac{\sigma E_\omega}{9\frac{D}{\omega R^2}} \sin 3 \theta,
\ee
and using Eq.~(\ref{firsteq}) to leading order in $\omega_c \tau$ gives
\be
\label{screeningresult}
I_{\omega^{<}}=-i  4\sigma E_\omega  \frac{\omega R^2}{9D} (\omega_c \tau)  R \int d\theta  \sin^2 3 \theta.
\ee
The imaginary factor implies that the AC current and pseudo-electric field $E_\omega$ are out of phase. Whereas $E_\omega \propto \omega$ so that in the high frequency regime $I_{\omega^>} \propto \omega$,  we see that in the low frequency regime $I_{\omega^<} \propto \omega^2$.

To estimate the dimensionless parameter $\frac{D}{\omega R^2}$, we use~\cite{sela2020quantum} the Einstein relation $\sigma  = D e^2 dn/d\mu$ and take $\sigma \sim e^2/h$ as the minimal conductivity of graphene~\cite{bolotin2008ultrahigh}. This yields 
\be
\frac{D}{\omega R^2} \sim \frac{v_F}{k_F \omega R^2}.
\ee
For $\omega=10^8$ Hz, $R=1 \mu$m, and $k_F = 10^2 \mu$m$^{-1}$, we have $\frac{D}{\omega R^2} \sim 100$. Thus, in the classical regime, Eq.~(\ref{screeningresult}) is applicable, rather than Eq.~(\ref{eq:19}).


Using Eqs.~(\ref{polarstraintensor}),~(\ref{magneticfield}), and (\ref{electricfield}), we estimate the maximal electric azimuthal field $(E_\omega)_{max}$ as
\be
(E_\omega)_{max} R \sim \frac{\omega R^2}{4}(B)_{max}\frac{\mathcal{U}^{AC}}{\mathcal{U}^{DC}} = \frac{1}{4} \omega (\delta B R^2).
\ee
Here $\delta B=B_{max}\frac{\mathcal{U}^{AC}}{\mathcal{U}^{DC}}$ is the amplitude of oscillation of the pseudo-magnetic field, and $(\delta B R^2)$ is of the order of the amplitude of oscillation of the magnetic flux. We thus obtain an estimate for the current
\bea
\label{estimateiomega}
|I_\omega| \cong \frac{2 \pi^2}{9} \frac{\omega_c \tau}{\frac{D}{\omega R^2}} \frac{\sigma}{\frac{e^2}{h}} (e \omega) \frac{\delta B R^2}{h/e} \cong 10 pA.
\eea
Here, $ (e \omega)$ has dimensions of current, and $\frac{\delta B R^2}{h/e}$ is the change in the number of flux quanta in the membrane in one period, suggesting a pumping mechanism~\cite{sabsovich2021helical}.   $\frac{\omega R^2}{D} $ is a suppression factor due to screening. In Eq.~(\ref{estimateiomega}) we used $B_{max}=1T$, $\frac{\mathcal{U}^{AC}}{\mathcal{U}^{DC}}=0.1$, and $\omega_c \tau=1/2$.
\subsection{Quantum regime}
Similar to Eq.~(\ref{6Landauer}), also the dynamic Hall effect has a quantum limit~\cite{sela2020quantum}. It takes the same form as Eq.~(\ref{6Landauer}),
\be
\label{6Landaueromega}
\frac{I_{\omega}}{V_\omega} = 6 \nu \frac{e^2}{h},
\ee
where now $V_\omega$ is the pseudo-voltage difference between $\theta=0$ and $\theta=\pi/3$ (mod $2 \pi /3$) zig-zag directions. See Sec.~\ref{appendix2} for a derivation. 
While the classical current is strongly suppressed in the  screening regime $\frac{D}{\omega R^2}\gg 1$, the quantum effect is unaffected by this factor, except at the Hall transitions~\cite{sela2020quantum}.
It also displays steps as function of filling factor $\nu$.

We estimate the current in this regime  as
\be
|I_\omega| \sim \frac{\pi \nu}{2} (\omega e) \frac{\delta B R^2}{h/e} \sim 1 nA.
\ee

The absence of screening in the quantum regime is understood via a pseudo-Landau level picture, which separates the edge states from the localized bulk states~\cite{sela2020quantum}. This is discussed in Sec.~\ref{appendix2}. The pseudo-electric field tilts the pseudo-Landau levels, leading to a non-decaying current.


\section{Pseudo-Landau levels and snake states for periodically varying field}
\label{appendix2}
In this section we solve the Dirac equation in an idealized geometry  in the presence of a periodically varying magnetic field. This allows  to visualize the pseudo-Landau levels (PLLs) and associated snake states. While we capture the angular dependence of the field, here we do not treat in detail its radial dependence.

\begin{figure}[h]
	\includegraphics[width=0.45\textwidth]{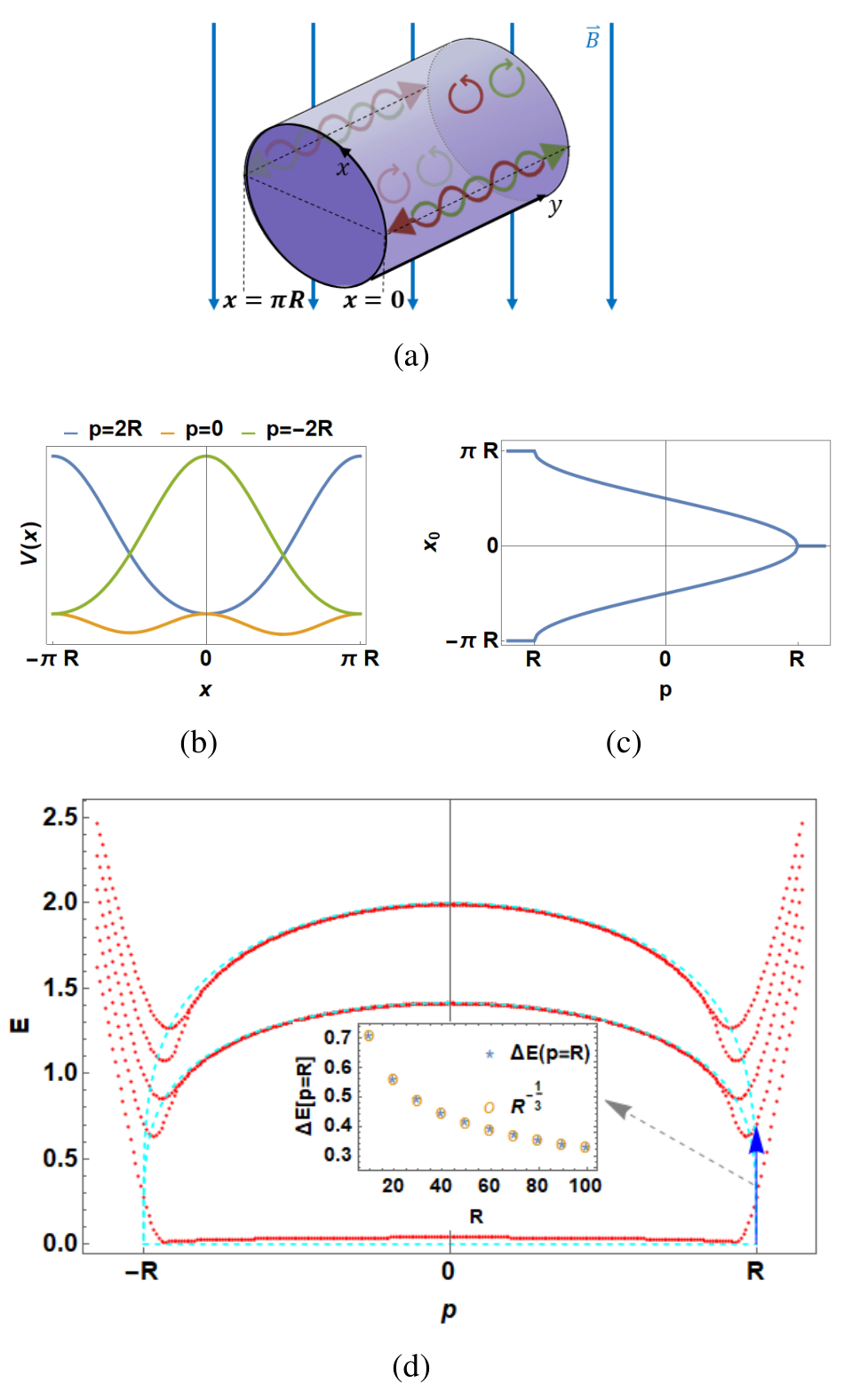}
	\caption{(Pseudo-)LLs on an infinite cylinder in a magnetic field. (a)  Illustration of semi-classical electronic trajectories on a cylinder (with angular coordinate $x$ and axial coordinate $y$) in the presence of a (pseudo-)magnetic field perpendicular to its axis. The two colors represent the two valleys. We depict a pair of counter-propagating snake states for $x=0$ and  $x=\pm \pi R$, at which the magnetic field perpendicular to the surface of the cylinder vanishes, as well as localized states otherwhere. (b) Potential profiles $V(x)$ in Eq.~(\ref{potential}) versus momentum $p$ (for large $R$). (c) Minima of $V(x)$ as function of $p$.  (d) Landau energy levels in magnetic units for valley $K$ [Eq.~(\ref{magneticfieldagunits})] computed numerically for $R=10$, and compared with Eqs.~(\ref{plls}) and (\ref{eq:bp}) (dashed). We  plot only the $n=0,1,2$ PLLs. For valley $K'$ the dispersion is unchanged but the relation between $p$ and $x$ is inverted.  Inset: scaling of the energy gap at $p=R$ as $1/R^{1/3}$.  }
	\label{fg:snakestates}
\end{figure}

The geometry we consider is an infinite cylinder with  angular coordinate  $x \in (0,2\pi R)$ and infinite axial coordinate $y$. This  is an idealization of the Corbino geometry, such that the radial coordinate $r \in (R-\Delta R/2,R+\Delta R/2)$ is treated as being infinite and translation invariant. This is reasonable when the  magnetic length $\ell_B=\sqrt{\frac{\hbar}{eB}}$ is much smaller than the dimensions of the system, $\ell_B \ll \Delta R,R$.
We add a magnetic field 
\be
\label{magneticfieldequal1}
B(x)=B\cos \frac{m x}{R},
\ee
with periodicity  $m=3$ as in Eq.~(\ref{magneticfield}). Below we  specialize to the case with $m=1$ which already displays the key features and the emergence of snake states; the generalization to $m=3$ is straightforward. In fact the $m=1$ magnetic field Eq.~(\ref{magneticfieldequal1}) has the physical interpretation of a cylinder embedded in a 3D space along a horizontal axis with a vertical magnetic field, see Fig.~\ref{fg:snakestates}(a).

Using the Landau gauge $A_x=0, A_y=-\frac{B R}{m}\cos \frac{mx}{R}$, we have the Dirac equation
\be
v_F[\sigma_x(-i \hbar \partial_x) +\sigma_y (p+\frac{eB R}{m})\cos \frac{mx}{R}  ]\Phi = E_p \Phi,
\ee
for the spinor $\Phi = (u(x),v(x))$ with $p$ the momentum along the infinite cylinder. By this equation we consider one valley, whereas the opposite valley has opposite field.

In the rest of  this section we define dimensionless coordinate, momentum, and energy variables 
\be
\label{magneticfieldagunits}
x \to x/\ell_B,~~~p \to p \ell_B/\hbar,~~~E_p \to E_p/(\hbar v_F/\ell_B).
\ee
We also replace $R \to R/\ell_B$. The resulting pair of equations is
\bea
O_-v&=&E_p u, \nonumber \\
O_+u&=&E_p v,
\eea
where
\be
O_{\pm}=-i\partial_x \pm i \left(p-\frac{R}{m} \cos\left( \frac{mx}{R}\right) \right).
\ee
This leads to the Schrodinger equation
\bea
\label{eq:SE}
O_+O_-v&=&E_p^2v,
\eea
where 
\bea
O_+ O_- =-\partial_x^2+\left(p-\frac{R}{m} \cos  \frac{mx}{R} \right)^2 - \sin  \frac{mx}{R}.
\eea
We now specify to $m=1$.  To gain physical understanding of the solutions we plot the potential, 
\be
V(x)=\left(p-R \cos  \frac{x}{R} \right)^2 + \sin  \frac{x}{R},
\ee
in Fig.~\ref{fg:snakestates}(b) for various $p$'s. Its minima  are shown in Fig.~\ref{fg:snakestates}(c)  as function of $p$. For  $p \gg R$ this potential has a minimum at $x_0=0$. 
As $p$  decreases from large positive values, the single minimum at $x_0=0$ splits at  $p \sim R$ into two minima which gradually separate from one another. At $p=0$  these minima reach approximately $\pm \pi R/2$ (with a height asymmetry due to the  $\sin(x/R)$ term in the potential, not shown) which then continuously move towards $\pm \pi R$ as $p$ further decreases. This doubly valued relation  
\be
\label{arccos}
x_0(p) \cong \pm R \arccos \frac{p}{R},
\ee
(neglecting the $\sin x/R$ term in $V(x)$) is used in Fig.~\ref{fg:snakestates}(c).

We show in Fig.~\ref{fg:snakestates}(a) schematic semiclassical solutions on the cylinder in a vertical magnetic field. Classically the solutions near $x_0=0,\pi R$, where $B$ changes sign, are snake states. Solutions for $0<x_0<\pi R$ display cyclotron motion clockwise and solutions for $0>x_0>-\pi R$ display cyclotron motion anti-clockwise. 
Next we construct the form of $E_p$ for the limit $R \gg 1$, namely the case when the dimensions of the system significantly exceed $\ell_B$. Expanding the potential around its minima, we have
\bea
V(x) & = & \left( p-R \cos  \frac{x_0+x-x_0}{R}  \right)^2-\sin   \frac{x_0+x-x_0}{R}   \nonumber \\
&\cong & b^2 (x-x_0)^2-b,
\eea 
where $b=b(x_0)=\sin \frac{x_0}{R}$. 
Thus 
\be
O_+O_- = -\partial_x^2 + b^2 (x-x_0)^2-b=b[-\frac{1}{b} \partial_x^2+b(x-x_0)^2 -1].
\ee
This harmonic oscillator  has eigenvalues 
\be
\label{plls}
E_p =\pm \sqrt{2 b n },
\ee
where $b=b(x_0(p))$. Using Eq.~(\ref{arccos}), 
\be
\label{eq:bp}
b  \cong \sqrt{1-(p/R)^2}.
\ee
This gives pseudo-Landau levels for $-R < p<R$ whose energy separation decreases towards the edges.   

For $|p| \gg R$, the $p-$dependence is captured by the expansion
\be
\label{potential}
V(x)=p^2-2pR \cos \frac{x}{R}\cong p^2,
\ee
with solutions $e^{i \frac{x}{ R}n}$
and hence $E_p = \pm \sqrt{ p^2+\frac{n^2}{R^2}}$. 

In Fig.~\ref{fg:snakestates}(d) we show the  energy spectrum obtained by numerical solution of Eq.~(\ref{eq:SE}) and compare it with the semiclassical result Eqs.~(\ref{plls}), (\ref{eq:bp}). Each Landau level is doubly degenerate (not taking into account spin and valley). This degeneracy corresponds to the upper and lower cyclotron solutions in the cylinder in Fig.~\ref{fg:snakestates}(a). We can see in Fig.~\ref{fg:snakestates}(d) how these doubly degenerate states split at $|p| \sim R$ and become edge states.

We see that the energy gap is minimal near $|p| \sim R$. As $|p|$ approaches $R$, the pair of wave functions at $\pm x_0$ approach each other, and their typical length $1/\sqrt{b}$ increases. Thus the $\pm x_0$ harmonic oscillator  solutions  will hybridize when $x \sim 1/\sqrt{b}$. This happens at $x \sim R^{1/3}$. The gap is then given by $E \sim R^{-1/3}$. This is justified using numerical calculation in the inset of Fig.~\ref{fg:snakestates}(d).


\subsection{Quantized conductance}
Having described the formation of (pseudo-)Landau levels, now we discuss the transport configurations in Fig.~1(c,d), in terms of our simplified cylinder. Along the way we will justify Eqs.~(\ref{6Landauer}) and (\ref{6Landaueromega}).

\begin{figure}[h]
	\includegraphics[width=0.45\textwidth]{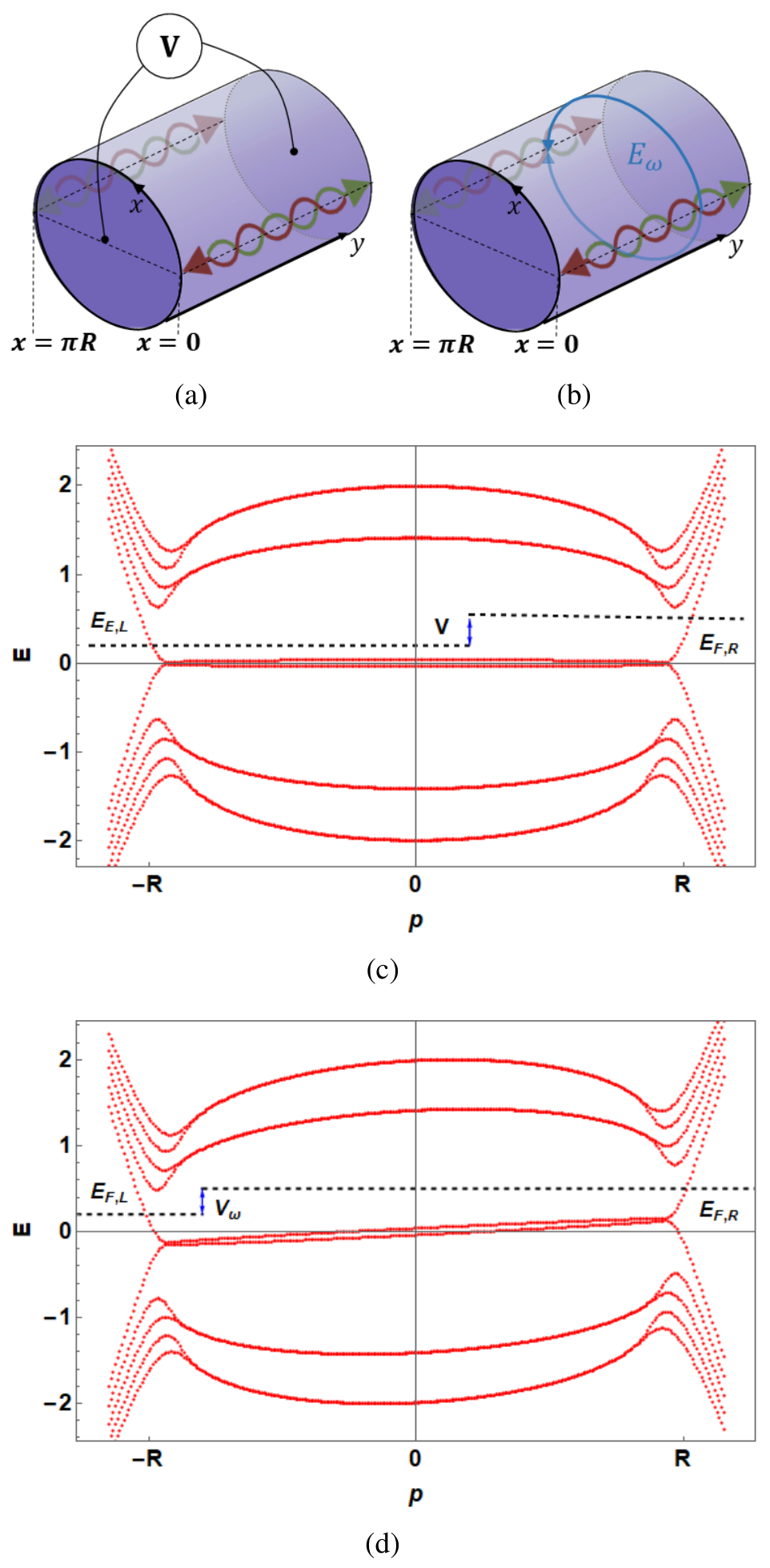}
	\caption{Schematics of pseudo-Landau levels under (a) an external voltage $V$ and (b) a pseudo-electric field leading to a pseudo-voltage $V_\omega$. In (c), the voltage $V$ sets the difference of chemical potential of the  counter propagating snake states. In (d), the internal (AC) pseudo-electric field leads to an (AC) potential Eq.~(\ref{addedv}) which tilts the pseudo-Landau levels, together with the corresponding Fermi energies. We used $V=V_\omega=0.3$ (in magnetic units). }
	\label{PLLstilted}
\end{figure}

Applying an external voltage $V$ on the Corbino geometry as in Fig.~1(c) corresponds to the cylinder of Fig.~\ref{PLLstilted}(a) connected between two contacts along $y$ at different chemical potentials. As a result we obtain a different Fermi energy $E_{F,R}$, $E_{F,L}$ for the snake states moving along the positive and negative $y$ directions, respectively, as shown in Fig.~\ref{PLLstilted}(c). In this schematic figure we assume that $\nu=2$ but other values can be considered. This leads to a quantized current
\be
\label{eqm1}
I = \frac{e}{h} \int_{occ} dp \frac{dE_p }{dp} =2 \nu \frac{e^2}{h} V~~~(m=1).
\ee
Here, $\nu$ is the number of chiral edge states that the Fermi levels cross at $p=R$ including spin $(\pm 2,\pm 6,\pm 10...)$ and the factor of 2 accounts for valley.

In the case with three-fold symmetry, $m=3$, each snake state is replaced by 3 such states, angularly shifted by $2\pi/3$, leading to Eq.~(\ref{6Landauer}).

Now we turn off the external voltage  $V$ and consider 
the configuration in Fig.~\ref{PLLstilted}(b) [corresponding to the dynamical case as in Fig.~1(d)]. The pseudo-electric field $\propto \sin \frac{m x}{R}$ adds a potential term $V_\omega(x) =- \int^x dx' E_\omega$. 
We take for illustration 
\be
\label{addedv}
V_\omega (x)= V_\omega \cos \frac{x}{R}.
\ee 
Assuming $R \gg 1$, we account for $V_\omega$  semi-classically by adding to each energy level $E_p$ a potential term $V_\omega(x_0)$ where $x_0$ is given in Eq.~(\ref{arccos}). The resulting pseudo-Landau levels tilt as in Fig.~1(d). Since the pseudo-electric field is produced dynamically, the tilt is also oscillatory. At each snapshot, the occupation is a non-equilibrium occupation with different Fermi energies at the counter propagating snake states, $E_{F,L/R}$. The occupation does not relax to the instantaneous ground state with a single Fermi energy because this would require diffusion through the localized bulk. The electronic current is given by Eq.~(\ref{eqm1}) for $m=1$, and similarly generalizes to Eq.~(\ref{6Landaueromega}) for $m=3$.


\section{Conclusions}
\label{sec:conclusions}
Strain fields in graphene are known to produce effective electromagnetic fields acting on the Dirac electrons. Here we analyzed a circularly symmetric Corbino membrane geometry as a path towards observing electronic transport through pseudo-magnetic and electric fields. 

In our system the pseudo-magnetic field is non-uniform and has a three-fold symmetry. In the static case, as discussed earlier~\cite{oroszlany2008theory,ghosh2008conductance,jones2017quantized}, the regions in which the magnetic field changes sign act as one-dimensional conductors carrying snake states, and may allow to observe quantized electronic transport. Time dependent strain obtained by AC gating generates an electronic current in the absence of an applied voltage. This dynamical pseudo-Hall effect persists although the pseudo-$E$ and $B$ fields are non-uniform and individually average to zero.

Therefore, the Corbino-disk membrane geometry is ideal to observe electronic transport in pseudo-magnetic and electric fields in general and snake states in particular. The quantized snake states, separated by localized bulk states, are essential to observe this dynamic current as well. While for a real magnetic field, the formation of snake states requires an antisymmetric field profile or an antisymmetric carrier
distribution~\cite{milovanovic2013spectroscopy,liu2015snake,rickhaus2015snake}, for the case of strain induced pseudo-fields, it is quite natural to have sign changes and even periodic variations of the pseudo-fields~\cite{banerjee2020strain}.  In our system, the magnetic length is smaller than the length over which the magnetic field changes sign, allowing to explore effects of the snake states in the quantum Hall regime.


\section{Acknowledgements} We acknowledge support by the US-Israel Binational
Science Foundation (Grant No. 2016255),  ARO (W911NF-20-1-0013), and the Israel Science Foundation grant number 154/19. We thank Moshe Ben Shalom, Kirill Bolotin, Benny Davidovitch, Paco Guinea, Daniel Sabsovich and Dan Klein for useful discussions.
\appendix
\section{Calculation of the pseudo-magnetic field  for large $R/\Delta R$}
\label{appendix1}
In this appendix we derive Eq.~(\ref{Banalitic}) for the pseudo-magnetic field in the narrow ring limit $\Delta R/R \ll 1$, and then test it against numerical simulations.

After solving Eq.~(\ref{diffuandhlim}) analytically, we notice that the magnetic field vanishes up to zero order of $\frac{1}{R}$.  To obtain the coefficient of $B\propto \frac{1}{R} $ we expand Eqs.~(\ref{diffuandh}) in the narrow ring limit, to one order in $\Delta R/R$ beyond Eq.~(\ref{diffuandhlim}). 
Doing this for the first Eq.~(\ref{diffuandh}) gives
\begin{align}
&u''=-h' h''-\frac{1}{R}\left(u'+ \frac{\mu}{\lambda+2\mu} h'^{2}\right).
\label{diffexpand}
\end{align}

Rather than solving this equation, we first expand magnetic field up to first order of $\frac{1}{R}$. Using Eq.~(\ref{magneticfield}) we have
\begin{align}
\mathcal{B}\left(\rho\right)=-u''-h'\cdot h'' +\frac{1}{R}h'^{2}+\frac{3}{R}u'.
\end{align}

Using Eq.~(\ref{diffexpand}), we get
\begin{align}
\mathcal {B}\left(\rho\right)&=\frac{1}{R} \left[4u'+\left( 1+\frac{\mu}{\lambda+2\mu} \right)h'^{2} \right]. 
\end{align}
Substituting Eqs.~(\ref{limitsol}), we finally obtain Eq.~(\ref{Banalitic}).

\begin{figure}[h]
	\includegraphics[width=0.45\textwidth]{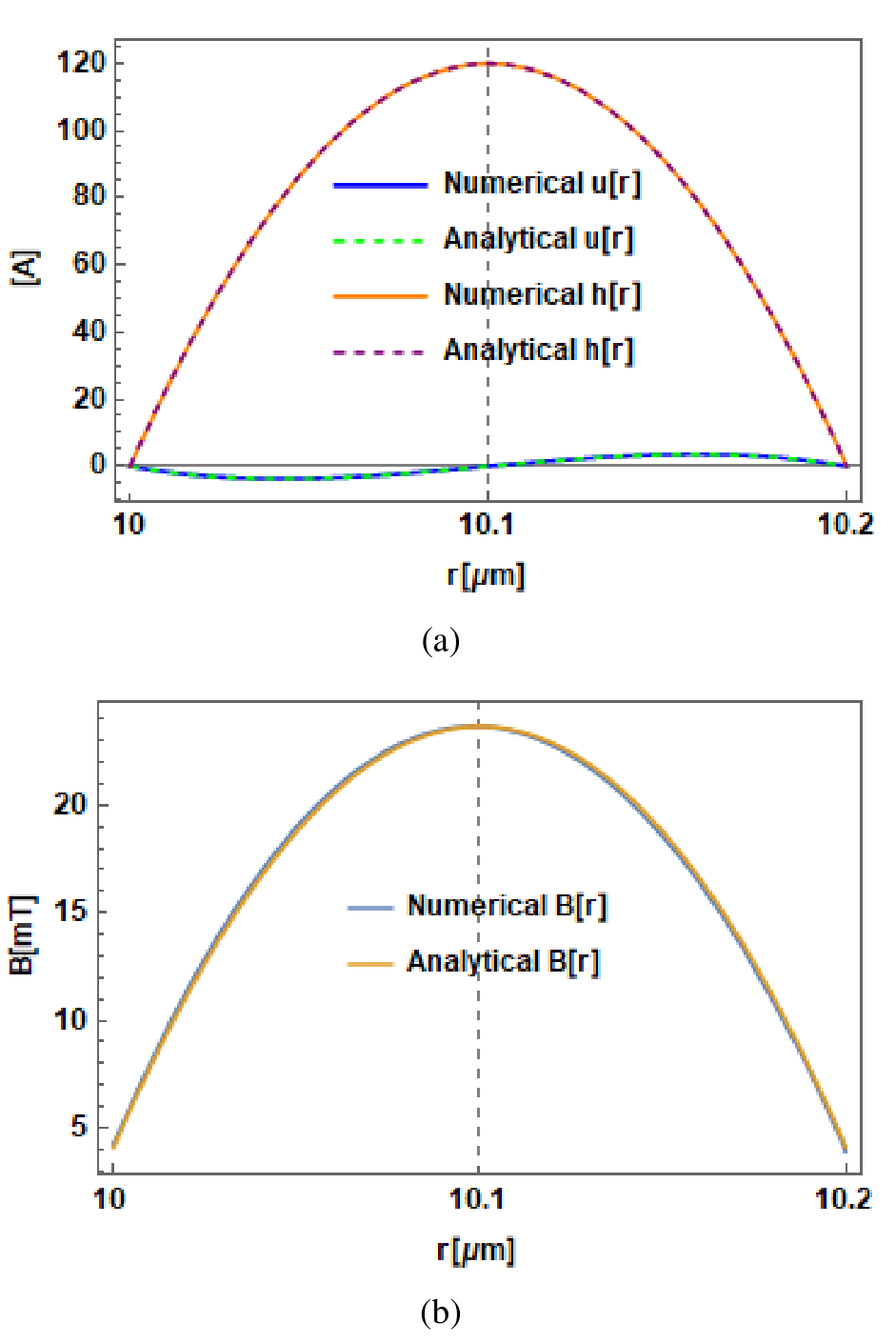}
	\caption{Narrow ring limit - comparison of numerical and analytical results. Top panel: Numerical solution of Eq.~(\ref{diffuandh}) for the in plane and out-of-plane displacements, compared to Eq.~(\ref{limitsol}). Bottom panel:  Resulting pseudo magnetic field $\mathcal{B}(r)$ obtained from Eq.~(\ref{magneticfield}) and compared with  Eq.~(\ref{Banalitic}). }
	\label{narrowringlimitg}
\end{figure}

We consider $R_{1}=10\mu m$ and $R_{2}=10.2 \mu m$, with $\Delta R=200nm \ll R$, allowing to compare with the narrow ring limit. The electron density is taken to be $n=7.5\cdot 10^{12}~{\rm{cm}}^{-2}$. The results shown in Fig.~\ref{narrowringlimitg} for the in plane displacement $u$, out of plane displacement $h$, and  pseudo magnetic field $\mathcal{B}(r)$ obtained from Eq.~(\ref{magneticfield}), agree with the analytic narrow ring limit. The pseudo-magnetic field is only of order of $10 {\rm{mT}}$ for the to large selected membrane dimension.

%

\end{document}